# Symmetry Breaking in BRST Quantization of SU(3)

Scott Chapman
Sunnyvale, CA

January, 2007

**Abstract**

New BRST-invariant states for SU(3) gauge field theory are presented. The states have finite norms and unlike the states that are usually used to derive path integrals, they break SU(3) symmetry by choosing preferred gauge directions. This symmetry breaking may also give effective masses to some of the gauge bosons of the theory.

## Introduction

The most common way to quantize SU(3) and other non-Abelian gauge theories is to use the path-integral formalism. In this formalism, path integrals represent the transition probability to go from one gauge-invariant state to another. In non-Abelian gauge theories, the nonlinear nature of the gauge constraints makes it very difficult to explicitly define and work with gauge-invariant states directly. By expanding the space of quantized fields to include ghosts, and by making a connection between gauge-invariance and BRST-invariance, the BRST method enables a straightforward definition of physical states that can be used in path integrals (see for example [1]). To avoid inconsistencies in the formulation, the underlying BRST-invariant states should also have finite norms [2].

The BRST-invariant states that are easiest to construct are those that are annihilated either by all of the gauge constraints (Dirac condition) or by all of the ghost operators. The main problem with using these states in path integrals is that they do not have well-defined norms. In a series of papers [2-7], it has been shown that if one multiplies these naïve BRST-invariant states by certain BRST-exact exponentials, one can create BRST-invariant states that do have finite norms. Using these finite-norm states, a consistent probabilistic interpretation is justified, and the path integral takes its usual form as the exponential of a gauge-constrained, classical action involving Fadeev-Popov ghosts.

In this paper, new finite-norm, BRST-invariant states are constructed for SU(3). Using these new states rather than the usual ones amounts to making a different choice for how to canonically quantize the underlying theory. Unlike the usual states that treat every gauge direction the same, the new states break the SU(3) symmetry by choosing preferred gauge directions. Although path integrals are not constructed from the new states, it may be that the action associated with these path integrals could develop effective mass terms for the gauge bosons.

The paper is organized as follows: In the first section, standard definitions are given, and the BRST-invariant states usually used in non-Abelian gauge theories are presented. In the second section, new notation is defined and the new solutions are presented. The last section contains summary comments and ideas for future work.

## Standard BRST Quantization

The calculations throughout the paper shall pertain to an SU(3) Yang-Mills gauge field theory, quantized using BRST canonical quantization. For convenience, space is assumed to be a three-dimensional grid with $N$ discreet points in a large volume $V$, so that integrals over space are replaced by sums over points, etc. The transition from discrete space back to continuous space will not be addressed in this paper.

The following definitions are used:

- Canonical fields, their momenta: $A_i^a, \Pi_i^a$
- Ghost fields, their momenta: $\eta_a, \mathcal{P}_a$
- Lagrange multipliers, their momenta: $\lambda_a, \pi_a$
- Anti-ghost fields, their momenta: $\overline{\eta}_a, \overline{\mathcal{P}}_a$
- Gauge constraints: $G_a = -\left(\partial_i \Pi_i^a - g f^{abc} \Pi_i^b A_i^c\right)$
- BRST charge: $\Omega = \sum_x \left[\eta_a G_a + \tfrac{1}{2} i g f^{abc} \eta_a \eta_b \mathcal{P}_c + \overline{\mathcal{P}}_a \pi_a\right]$
- Ghost number operator: $\mathcal{G} = \tfrac{1}{2} \sum_x \left( [\eta_a, \mathcal{P}_a] - [\overline{\eta}_a, \overline{\mathcal{P}}_a] \right),$ (1)

All of the operators defined above are Hermitian except for the ghost number operator, which is anti-Hermitian. $f^{abc}$ is the totally antisymmetric SU(3) structure constant, and the spatial derivatives $\partial_i$ in the gauge constraints are defined as differences in the usual way for discrete-space analyses.

Canonical quantization is achieved through the following commutators and anticommutators:

$$[\Pi_i^a(\vec{x},t), A_j^b(\vec{y},t)] = -i\delta_{ij}\delta^{ab}\delta_{xy}$$

$$[\pi_a(\vec{x},t), \lambda_b(\vec{y},t)] = -i\delta_{ab}\delta_{xy}$$

$$\{\eta_a(\vec{x},t), \mathcal{P}_b(\vec{y},t)\} = \delta_{ab}\delta_{xy}$$

$$\{\overline{\eta}_a(\vec{x},t), \overline{\mathcal{P}}_b(\vec{y},t)\} = \delta_{ab}\delta_{xy} \qquad (2)$$

Since calculations are being made in discrete space rather than continuous space, Kroenecker delta functions are used rather than Dirac delta functions. Using these commutation relations, it is straightforward to see that the gauge constraints close in an SU(3) group under commutation,

$$[G_a(\vec{x},t), G_b(\vec{y},t)] = -ig\delta_{xy}f^{abc}G_c(\vec{x},t), \qquad (3)$$

and that the BRST charge is nilpotent, $\Omega^2 = 0$.

In BRST quantization, the physical states of the theory must be annihilated by the BRST charge. For non-Abelian theories, this is most easily accomplished by imposing one of the following two conditions on physical states:

$$\pi_a|\tilde{\xi}_\alpha\rangle = \eta_a|\tilde{\xi}_\alpha\rangle = \overline{\eta}_a|\tilde{\xi}_\alpha\rangle = 0 \qquad \text{or} \qquad (4a)$$

$$G_a|\tilde{\xi}'_\alpha\rangle = \mathcal{P}_a|\tilde{\xi}'_\alpha\rangle = \overline{\mathcal{P}}_a|\tilde{\xi}'_\alpha\rangle = 0. \qquad (4b)$$

where the index "$\alpha$" specifies the field structure of a particular state which meets one of the above two conditions. Both of the states above are annihilated by the ghost number operator $\mathcal{G}$, so they are ghost number eigenstates with ghost number equal to zero.

A problem with the states in (4) is that they do not have well-defined norms. For example, after expanding the states in (4a) in the Schroedinger representation (and ghost momentum representation), one finds:

$$\langle\tilde{\xi}_\alpha\|\tilde{\xi}_\alpha\rangle = \int\prod_x\left[dA_i^a d\lambda_a d\mathcal{P}_a d\overline{\mathcal{P}}_a|\psi_\alpha(A_i^a(\vec{x}))|^2\right] \qquad (5)$$

where $\psi_\alpha(A_i^a(\vec{x})) = \langle A_i^a(\vec{x})\|\tilde{\xi}_\alpha\rangle$ is the wave function of the physical state with index $\alpha$. Since $\pi_a|\tilde{\xi}_\alpha\rangle = 0$, the wave functions are independent of $\lambda_a$, so integrations over the Lagrange multipliers produce factors of infinity. At the same time, the Berezin integrals over $d\mathcal{P}_a$ and $d\overline{\mathcal{P}}_a$ are equal to zero. Thus the states defined by the condition (4a) have ill-defined norms. Similar arguments apply to the states defined by (4b).

In [4], it was pointed out that BRST-invariant states with finite norms can be constructed if one multiplies these states by BRST-exact exponentials and quantizes the Lagrange-multipler sector with an indefinite metric. Consider the states:

$$|\xi_\alpha\rangle = \exp\left(\left\{\Omega,-\varepsilon\sum_x \lambda_a \mathcal{P}_a\right\}\right)|\tilde{\xi}_\alpha\rangle = \exp\left(-\varepsilon\sum_x \left(\lambda_a \tilde{G}_a - i\overline{\mathcal{P}}_a \mathcal{P}_a\right)\right)|\tilde{\xi}_\alpha\rangle, \qquad (6)$$

where $\varepsilon$ is a real, nonzero constant and

$$\tilde{G}_a = G_a + igf^{abc}\eta_b \mathcal{P}_c. \qquad (7)$$

Since the exponent of (6) is a BRST-exact function, those states are still BRST-invariant. They also still have zero ghost number. Furthermore, they have well-defined norms given by:

$$\langle\xi_\alpha\|\xi_\alpha\rangle = \int \prod_{x,a}\left[d\mathcal{P}_a \mathcal{P}_a d\overline{\mathcal{P}}_a \overline{\mathcal{P}}_a dA_i^a \,\delta(G_a)|\psi_\alpha(A_i^a(\vec{x}))|^2\right] \qquad (8)$$

that are independent of $\varepsilon$. In the above expression, the Berezin integrals over the ghosts and anti-ghosts no longer vanish, but produce factors of 1. At the same time, integrations over $\lambda_a$ no longer produce infinities, but instead produce the delta functions $\delta(G_a)$. A subtle requirement in producing these delta functions is that the Lagrange-multipler momentum states $|\lambda_a\rangle$ must be quantized with indefinite metric. In the context of this quantization, the Hermitian operators $\lambda_a$ have imaginary eigenvalues [1,8], leading to delta functions upon integration.

Path integrals can be derived by using the states in (6) to calculate projected kernels of the operator $\exp(i\varepsilon H)$, where $H$ is the Hamiltonian and $\varepsilon$ is an infinitesimal time interval. The process of projecting the kernels onto the physical states of (6) effectively adds a term to the Hamiltonian:

$$H \to H + \sum_x \left(A_0^a \tilde{G}_a + \overline{\mathcal{P}}_a \mathcal{P}_a\right) \qquad (9)$$

where $A_0^a \equiv -i\lambda_a$ is an anti-Hermitian operator with real eigenvalues (due to the indefinite metric). By inserting a complete set of canonical momentum states, integrating over the momenta, and going to the continuum limit, the path integral kernel takes its usual form as the exponential of a gauge-constrained, classical action involving Fadeev-Popov ghosts. Later in the paper, it will be seen that the new solutions presented here lead to differences in the path integral.

In any case, a prerequisite to developing a consistent path integral formulation is to start with BRST-invariant states that have finite norms.

There is another way to prove that the states in (6) have finite norms. In [2,3], it was shown that if BRST-invariant states are independently annihilated by two conjugate, nilpotent operators whose sum is equal to the BRST charge, then those states have finite norms, provided that they satisfy the following additional conditions. The nilpotent operators must have a ghost factor $c_a$ and a constraint factor $\phi_a$ (which may also contain ghosts), and if the constraint annihilates a state, then the conjugate of the constraint should not annihilate the same state. These requirements can be summed up by:

$$\Omega = \delta + \delta^\dagger$$
$$\delta^2 = 0$$
$$\delta = c_a^\dagger \phi_a$$
$$c_a |\xi_\alpha\rangle = \phi_a |\xi_\alpha\rangle = \mathcal{G} |\xi_\alpha\rangle = 0$$
$$f(\phi_a^\dagger)|\xi_\alpha\rangle \neq 0 \quad \text{unless} \quad f = 0. \tag{9}$$

The last "canonical" condition allows one to represent the matter states canonically in such a way that the gauge constraint degrees of freedom can be factored out. In [4], it was shown that the states of eq. (6) satisfy these conditions and are therefore BRST-invariant states with finite norms.

## New classes of BRST-invariant states for SU(3)

In this section, new BRST-invariant states will be explicitly constructed. Unlike the states considered in the last section, these new states do not mix the minimal and non-minimal (Lagrange multiplier) sectors for the off-diagonal gauge directions. As a result, in the context of these states, the off-diagonal non-minimal sector decouples, so one need only consider a BRST charge with non-minimal contributions in the diagonal gauge directions:

$$\Omega' = \sum_x \left[ \eta_a G_a + \tfrac{1}{2} i g f^{abc} \eta_a \eta_b \mathcal{P}_c + \overline{\mathcal{P}}_3 \pi_3 + \overline{\mathcal{P}}_8 \pi_8 \right] . \tag{10}$$

Using this BRST charge, the states presented in this section will be shown to be BRST-invariant with finite norms by satisfying the conditions of (9).

State construction begins with the following definitions which preserve ghost number:

$$\eta_1^\pm = (\eta_1 \mp i\eta_2) \qquad \mathcal{P}_1^\pm = \tfrac{1}{2}(\mathcal{P}_1 \mp i\mathcal{P}_2) \qquad G_1^\pm = \tfrac{1}{2}(G_1 \mp iG_2)$$
$$\eta_2^\pm = (\eta_4 \pm i\eta_5) \qquad \mathcal{P}_2^\pm = \tfrac{1}{2}(\mathcal{P}_4 \pm i\mathcal{P}_5) \qquad G_2^\pm = \tfrac{1}{2}(G_4 \pm iG_5)$$

$$\eta_3^\pm = (\eta_6 \mp i\eta_7) \qquad \mathcal{P}_3^\pm = \tfrac{1}{2}(\mathcal{P}_6 \mp i\mathcal{P}_7) \qquad G_3^\pm = \tfrac{1}{2}(G_6 \mp iG_7)$$

$$c_4^\pm = (\eta_3 \mp i\varepsilon\overline{\mathcal{P}}_3) \qquad k_4^\pm = \tfrac{1}{2}\left(\mathcal{P}_3 \mp i\tfrac{1}{\varepsilon}\overline{\eta}_3\right) \qquad \psi_4^\pm = \tfrac{1}{2}\left(G_3 \mp i\tfrac{1}{\varepsilon}\pi_3\right)$$

$$c_5^\pm = (\eta_8 \mp i\varepsilon\overline{\mathcal{P}}_8) \qquad k_5^\pm = \tfrac{1}{2}\left(\mathcal{P}_8 \mp i\tfrac{1}{\varepsilon}\overline{\eta}_8\right) \qquad \psi_5^\pm = \tfrac{1}{2}\left(G_8 \mp i\tfrac{1}{\varepsilon}\pi_8\right) , \qquad (11)$$

where $\varepsilon$ is a nonzero constant. The indices on the ghost fields on the right-hand sides of the above equations assume a basis for SU(3) defined by the standard 8 Gell-Mann 3x3 matrices. It should be noted that the operators in the Gell-Mann "5" direction have different signs from those in the "2" and "7" directions, and that the third "off-diagonal" constraints $G_3^\pm$ are distinct from the constraint $G_3$ that is in the diagonal "3" gauge direction.

The BRST charge can be rewritten in terms of the operators in (11). One finds:

$$\Omega' = \sum_{\vec{x}} [\Omega_1(\vec{x}) + \Omega_2(\vec{x}) + \Omega_3(\vec{x})]$$

$$\Omega_1 = \eta_i^- G_i^+ + \eta_i^+ G_i^- - \tfrac{1}{4} g \varepsilon^{ijk}\left(\eta_i^+ \eta_j^+ \mathcal{P}_k^+ + \eta_j^- \eta_i^- \mathcal{P}_k^-\right)$$

$$\Omega_2 = \tfrac{1}{4} g (2\eta_1^+\eta_1^- - \eta_2^+\eta_2^- - \eta_3^+\eta_3^-)(k_4^+ + k_4^-) + \tfrac{1}{4} g\sqrt{3}(-\eta_2^+\eta_2^- + \eta_3^+\eta_3^-)(k_5^+ + k_5^-)$$

$$\Omega_3 = c_4^- \psi_4^+ + c_4^+ \psi_4^- + c_5^- \psi_5^+ + c_5^+ \psi_5^- + \tfrac{1}{4} g(c_4^+ + c_4^-)N_3 + \tfrac{1}{4} g\sqrt{3}(c_5^+ + c_5^-)N_8$$

$$N_3 \equiv 2\eta_1^+\mathcal{P}_1^- - 2\eta_1^-\mathcal{P}_1^+ - \eta_2^+\mathcal{P}_2^- + \eta_2^-\mathcal{P}_2^+ - \eta_3^+\mathcal{P}_3^- + \eta_3^-\mathcal{P}_3^+$$

$$N_8 \equiv -\eta_2^+\mathcal{P}_2^- + \eta_2^-\mathcal{P}_2^+ + \eta_3^+\mathcal{P}_3^- - \eta_3^-\mathcal{P}_3^+ , \qquad (12)$$

where the indices *i, j, k* run over the values 1, 2, 3. This form of the BRST charge highlights the threefold symmetry of the "off-diagonal" generators of SU(3), especially in the term in $\Omega_1$ that has a factor of $\varepsilon^{ijk}$.

A prerequisite to satisfying eqs (9) is that one must be able to split the gauge constraints into two conjugate groups that separately close under commutation. Consider the following groupings of gauge constraints:

$$_1\phi_1^- \equiv G_1^+ \qquad\qquad _3\phi_1^- \equiv G_1^- \qquad\qquad _5\phi_1^- \equiv G_1^-$$

$$_1\phi_2^- \equiv G_2^- \qquad\qquad _3\phi_2^- \equiv G_2^+ \qquad\qquad _5\phi_2^- \equiv G_2^-$$

$$_1\phi_3^- \equiv G_3^- \qquad\qquad _3\phi_3^- \equiv G_3^- \qquad\qquad _5\phi_3^- \equiv G_3^+$$

$$_1\phi_4^- \equiv \psi_4^- + g \qquad\quad _3\phi_4^- \equiv \psi_4^- - \tfrac{1}{2}g \qquad\quad _5\phi_4^- \equiv \psi_4^- - \tfrac{1}{2}g$$

$$_1\phi_5^- \equiv \psi_5^- \qquad\qquad _3\phi_5^- \equiv \psi_5^- - \frac{\sqrt{3}}{2}g \qquad _5\phi_5^- \equiv \psi_5^- + \frac{\sqrt{3}}{2}g \qquad (13)$$

The five constraints in each column close under commutation, while their complex conjugates separately close. Each grouping will be called a "class", labeled with a left lower index. The use of only odd numbering of classes in (13) reflects the fact that three additional classes can be constructed as follows: On the right hand sides of the above equations, change the sign of the constant terms in the fourth and fifth constraints, and change all of the – indices to + indices and vice versa. The new classes created in this way from classes 1, 3, and 5 will be numbered 2, 4, and 6, respectively.

For each of the classes above, one can separate the BRST charge into two separately nilpotent, conjugate components. For example, for the first class above, one can define:

$$\delta_1 = \eta_1^- G_1^+ + \eta_2^+ G_2^- + \eta_3^+ G_3^- - \tfrac{1}{4} g \varepsilon^{ijk} \eta_i^+ \eta_j^+ \mathcal{P}_k^+ + c_4^+ \psi_4^- + c_5^+ \psi_5^- + \tfrac{1}{4} g c_4^+ N_3 + \tfrac{1}{4} g \sqrt{3} c_5^+ N_8$$
$$+ \tfrac{1}{4} g \left( 2\eta_1^+ \eta_1^- - \eta_2^+ \eta_2^- - \eta_3^+ \eta_3^- \right) k_4^+ + \tfrac{1}{4} g \sqrt{3} \left( -\eta_2^+ \eta_2^- + \eta_3^+ \eta_3^- \right) k_5^+ . \qquad (14)$$

It is straightforward to verify that the above operator is nilpotent and satisfies $\Omega' = \delta_1 + \delta_1^\dagger$. Similar separations can be made for the other two classes of (13). Separations similar to that of (14) were pointed out in [3] for more general Lie groups.

For each class in (13) one can also define "matter" states $\left| _n M_\alpha \right\rangle$ that satisfy

$$_n\phi_\mu^- \left| _n M_\alpha \right\rangle = 0 \qquad \text{but} \qquad _n\phi_\mu^+ \left| _n M_\alpha \right\rangle \neq 0 . \qquad (15)$$

In the above equations, the index $n$ denotes one of the classes of eq (13), the index $\mu$ runs over the 5 constraints in each class, and the index $\alpha$ runs over all of the matter states that satisfy the above conditions. It has been shown in [2,3] that matter states that satisfy these conditions are canonical with finite inner products, so one may take them to be orthonormal:

$$\left\langle _n M_\alpha \| _n M_\beta \right\rangle = \delta_{\alpha\beta} . \qquad (16)$$

The ghost states for each class can be specified as follows:

$$\eta_1^+ \left| _1\mathcal{G} \right\rangle = \mathcal{P}_1^+ \left| _1\mathcal{G} \right\rangle = \eta_2^- \left| _1\mathcal{G} \right\rangle = \mathcal{P}_2^- \left| _1\mathcal{G} \right\rangle = \eta_3^- \left| _1\mathcal{G} \right\rangle = \mathcal{P}_3^- \left| _1\mathcal{G} \right\rangle = 0$$
$$\eta_1^- \left| _3\mathcal{G} \right\rangle = \mathcal{P}_1^- \left| _3\mathcal{G} \right\rangle = \eta_2^+ \left| _3\mathcal{G} \right\rangle = \mathcal{P}_2^+ \left| _3\mathcal{G} \right\rangle = \eta_3^- \left| _3\mathcal{G} \right\rangle = \mathcal{P}_3^- \left| _3\mathcal{G} \right\rangle = 0$$
$$\eta_1^- \left| _5\mathcal{G} \right\rangle = \mathcal{P}_1^- \left| _5\mathcal{G} \right\rangle = \eta_2^- \left| _5\mathcal{G} \right\rangle = \mathcal{P}_2^- \left| _5\mathcal{G} \right\rangle = \eta_3^+ \left| _5\mathcal{G} \right\rangle = \mathcal{P}_3^+ \left| _5\mathcal{G} \right\rangle = 0$$
$$c_k^- \left| _n\mathcal{G} \right\rangle = k_k^- \left| _n\mathcal{G} \right\rangle = 0 , \qquad (17)$$

where the left lower index on the ghost states is again being used to identify the class. The even number classes 2, 4, and 6 are found from 1, 3, and 5 by changing the sign of the upper indices in the first three lines of (17). All of the ghost states have vanishing ghost number, and they can each be assumed to have a norm equal to 1.

With the above definitions, the following states are easily verified to be BRST-invariant:

$$\left|{}_n\xi_\alpha\right\rangle = \left|{}_n\mathcal{G}\right\rangle\left|{}_nM_\alpha\right\rangle \tag{18}$$

Furthermore, due to (16), the states in any give class are orthonormal. It is interesting that each of the states breaks the threefold symmetry of SU(3) by treating one off-diagonal direction differently from the other two.

Another interesting feature of these states can be seen by writing their diagonal sectors in a form similar to that of eq. (6). For example, one may rewrite $\left|{}_1\xi_\alpha\right\rangle$ as follows:

$$\left|{}_1\xi_\alpha\right\rangle = \exp\left(-\varepsilon\left\{\Omega, K_1\right\}\right)\left|{}_1\widetilde{\mathcal{G}}\right\rangle\left|{}_1\widetilde{M}_\alpha\right\rangle$$

$$\left\{\Omega, K_1\right\} = \sum_x \lambda_3\left(G_3 + \tfrac{1}{2}gN_3\right) + \lambda_8\left(G_8 + \tfrac{1}{2}\sqrt{3}N_8\right) - i\overline{\mathcal{P}}_3\mathcal{P}_3 - i\overline{\mathcal{P}}_8\mathcal{P}_8 \tag{19}$$

In order to ensure that (15) and (17) continue to hold, these new matter and ghost states become Schroedinger-like in the diagonal sector (compare to (4a)),

$${}_1\phi_i^-\left|{}_1\widetilde{M}_\alpha\right\rangle = \pi_3\left|{}_1\widetilde{M}_\alpha\right\rangle = \pi_8\left|{}_1\widetilde{M}_\alpha\right\rangle = 0$$

$$\eta_3\left|{}_1\widetilde{\mathcal{G}}\right\rangle = \eta_8\left|{}_1\widetilde{\mathcal{G}}\right\rangle = \overline{\eta}_3\left|{}_1\widetilde{\mathcal{G}}\right\rangle = \overline{\eta}_8\left|{}_1\widetilde{\mathcal{G}}\right\rangle = 0, \tag{20}$$

and the ghost states also still satisfy the conditions in the first line of (17). The exponent in (19) simplifies due to the fact that each of the ghost states in (13) are eigenstates of the operators $N_3$ and $N_8$. One finds

$$\left\{\Omega, K_1\right\}\left|{}_1\widetilde{\mathcal{G}}\right\rangle = \sum_x \left[\lambda_3\left(G_3 + 2g\right) + \lambda_8 G_8 - i\overline{\mathcal{P}}_3\mathcal{P}_3 - i\overline{\mathcal{P}}_8\mathcal{P}_8\right]\left|{}_1\widetilde{\mathcal{G}}\right\rangle. \tag{21}$$

The most surprising thing about the above operator is that it contains a term that is linear in $\lambda_3$.

To understand the implications of this term, note that the Lagrange-multipler states $\left|\lambda_a\right\rangle$ must again be quantized with indefinite metric in order for the integrations over $\lambda_3$ and $\lambda_8$ to produce delta functions $\delta(G_3 + 2g)$ and $\delta(G_8)$ which are consistent with the constants in the fourth and fifth lines of (13). Due to this indefinite metric, one can again define $A_0^a \equiv -i\lambda_a$, where $A_0^a$ is an anti-Hermitian operator with real eigenvalues. In the path integral, one can then make a replacement similar to that of (9),

$$H \to H + \sum_x \left[ A_0^3 (G_3 + 2g) + A_0^8 G_8 + \overline{\mathcal{P}}_3 \mathcal{P}_3 + \overline{\mathcal{P}}_8 \mathcal{P}_8 \right] , \qquad (22)$$

but this time there is a term linear in $A_0^3$. This linear term suggests that $A_0^3$ could develop a vacuum expectation value (VEV).

To correctly determine whether or not a VEV does in fact develop and to understand its implications, it would be necessary to first consistently develop the path integral formulation for the hybrid states considered here. Although the states in (20) are Schroedinger-like in the diagonal sector, they are Fock-like in the off-diagonal sector. Construction of the path integral formalism for these states is outside the scope of this paper. However, by drawing parallels with more standard path integral approaches, one may find some clues as to what a VEV for $A_0^3$ could imply.

In standard path integral approaches, generated for example by states like those in (6), after one inserts a complete set of momentum states and integrates over the momentum, the resulting classical action in the exponent has "electric" terms in it like

$$\tfrac{1}{2}\left(F_{0i}^1\right)^2 = \tfrac{1}{2}\left(\partial_0 A_i^1 - \partial_i A_0^1 - g A_0^3 A_i^2 + ...\right)^2 . \qquad (23)$$

If $A_0^3$ were to develop a VEV, terms like those above would generate negative mass-squared terms for the off-diagonal gauge fields. These negative mass-squared terms could cause some spatial components of the off-diagonal gauge fields to also develop VEVs. Those spatial VEVs would generate positive mass-squared terms that would at least offset the negative mass-squared terms caused by $A_0^3$. The net result of this symmetry breaking and formation of VEVs could be effective masses for several of the gauge bosons.

## Summary

Six new classes of BRST-invariant states with finite norms have been explicitly constructed for SU(3) gauge field theory. The new classes represent different ways to canonically quantize SU(3), so only one class can exist in any particular region of space. It may be possible to make different quantization choices in different regions of space, and it would be interesting to explore the physical implications of such a program, particularly to see if they would shed more light on the mechanism behind QCD confinement.

Each of these new classes of states also breaks the SU(3) symmetry in some way. As a consequence of this symmetry breaking, path integrals derived from the states feature new terms in the Hamiltonian that are linear in $A_0^3$ and $A_0^8$. It would be interesting to fully develop the path integral formulation in the context of these states to see if those linear terms would produce gauge boson masses as they would in standard path integrals.